\documentclass[aps,notitlepage, floatfix, prb, twocolumn]{revtex4-2}

\usepackage[colorlinks=true, allcolors=blue, bookmarks=false]{hyperref}
\usepackage{amsfonts, amsmath, amssymb}
\usepackage{graphicx}

\DeclareMathOperator{\diag}{diag}

\begin{document}

\title{Long-range spectral statistics of the Rosenzweig-Porter model}

\author{Wouter Buijsman} 

\email{bwouter@pks.mpg.de}

\affiliation{Department of Physics, Ben-Gurion University of the Negev, Beer-Sheva 84105, Israel}

\affiliation{Max Planck Institute for the Physics of Complex Systems, N\"othnitzer Str. 38, 01187 Dresden, Germany}

\date{\today}

\begin{abstract}
The Rosenzweig-Porter model is a single-parameter random matrix ensemble that supports an ergodic, fractal, and localized phase. The names of these phases refer to the properties of the (midspectrum) eigenstates. This work focuses on the long-range spectral statistics of the recently introduced unitary equivalent of this model. By numerically studying the Thouless time obtained from the spectral form factor, it is argued that long-range spectral statistics can be used to probe the transition between the ergodic and the fractal phases. The scaling of the Thouless time as a function of the model parameters is found to be similar to the scaling of the spreading width of the eigenstates. Provided that the transition between the fractal and the localized phases can be probed through short-range level statistics, such as the average ratio of consecutive level spacings, this work establishes that spectral statistics are sufficient to probe both transitions present in the phase diagram. 
\end{abstract}

\maketitle

\section{Introduction}
Spectral (level) statistics provide a convenient, basis-independent probe for quantum chaos \cite{Porter65, Brody81, Mehta91}. Starting from early studies on the spectra of heavy atomic nuclei, spectral statistics are nowadays frequently used to characterize phases of matter in, for example, studies on single-body (Anderson) \cite{Shklovskii93, Mirlin00, Evers08, Suntajs23} and many-body \cite{Oganesyan07, Serbyn16, Buijsman19, Sierant20, Sierant20-2, Suntajs20} localization, random matrix theory \cite{Gharibyan18, Vleeshouwers21, Vleeshouwers22}, and integrability \cite{Caux11, Szasz-Schagrin21}. Broadly speaking, spectral statistics come in two types: short and long range. Short-range spectral statistics are commonly quantified by the level spacing distribution or the average ratio of consecutive level spacings \cite{Atas13, Giraud22}, while long-range spectral statistics are typically studied by focusing on the spectral form factor \cite{Kos18, Cipolloni23}, next to others \cite{Mehta91}. Often, studies on short- and long-range spectral statistics provide complementary results.

The Rosenzweig-Porter model is a single-parameter random matrix ensemble that supports an ergodic, fractal (also known as delocalized yet nonergodic), and localized phase \cite{Rosenzweig60, Kravtsov15}. These names refer to the fractal properties of the (mid-spectrum) eigenstates. In the thermodynamic limit, the phase diagram of this model can be obtained fully by analytical methods \cite{Facoetti16, Truong16, Monthus17, Bogomolny18, VonSoosten19}. Particularly well-studied from an analytical point of view are the two-point spectral correlations at the transition between the fractal and localized phases \cite{Pandey95, Brezin96, Altland97, Kunz98}. The Rosenzweig-Porter model serves as a natural toy model for the many-body localization transition as there is a fractal phase in-between the ergodic and localized phases, similar to what has been observed in physical models \cite{Mace19, Luitz20}. Recently, part of the phase diagram of this model has been observed experimentally \cite{Zhang23}. During the last few years, various variants and generalizations of the Rosenzweig-Porter model, for example with multifractal eigenstates, have been proposed \cite{DeTomasi19, Khaymovich20, Biroli21, Khaymovich21, Kutlin21, Buijsman22, DeTomasi22, Venturelli23, Barney23, Sarkar23, Roy23, Kutlin24}.

Long-range spectral statistics are studied numerically most conveniently for unitary models, for which the eigenvalues are located on the unit circle in the complex plane. Such models typically have a uniform density of states, meaning that no unfolding (uniformizing the density of states) is required. The absence of spectral edges then also guarantees that results are not affected by deviating edge statistics. A unitary equivalent of the (Hermitian) Rosenzweig-Porter model has been proposed recently in Ref. \cite{Buijsman22}. The unitary model can, for example, be used as a toy model for the many-body localization transition in periodically driven (Floquet) systems \cite{Ponte15, Abanin16, Zhang16, Bordia17}. Complementing recent interest in the spectral statistics of the Rosenzweig-Porter model and its variants \cite{Berkovits20, Venturelli23}, this work focuses on long-range spectral statistics of the unitary equivalent of the Rosenzweig-Porter model. By numerically studying the Thouless time obtained from the spectral form factor, it is argued that long-range spectral statistics can be used to probe the transition between the ergodic and the fractal phases. The scaling of the Thouless time as a function of the model parameters is found to be similar to the scaling of the (inverse) spreading width of the eigenstates. Provided that the transition between the fractal and the localized phases can be probed through short-range spectral statistics, this work establishes that spectral statistics are sufficient to probe both transitions present in the phase diagram.

The outline of this work is as follows. Section \ref{sec: Rosenzweig-Porter} introduces the Rosenzweig-Porter model and the unitary equivalent of it. Section \ref{sec: short-range} introduces the probes and discusses the results for short-range spectral (level spacing) statistics. Section \ref{sec: long-range}, which contains the main results, introduces the probes and discusses the results for long-range spectral statistics. Section \ref{sec: conclusions} concludes with a summary and outlook.

\section{Rosenzweig-Porter model and its unitary equivalent} \label{sec: Rosenzweig-Porter}
The (Hermitian) Rosenzweig-Porter model with complex-valued elements consists of matrices $H$ of the form
\begin{equation}
H = H_0 +\frac{1}{\sqrt{N^\gamma}} \, M.
\end{equation}
Here, $N$ is the matrix dimension, and $\gamma \ge 0$ is a tuning parameter. The matrix $H_0$ is diagonal with the non-zero elements sampled independently from the normal distribution with mean zero and unit variance. The matrix $M$ is a sample from the Gaussian unitary random matrix ensemble. An $N \times N$ matrix $M$ sampled from this ensemble can be constructed as
\begin{equation}
M = \frac{1}{2} ( A + A^\dagger),
\end{equation}
where $A$ is an $N \times N$ matrix with complex-valued elements $A_{nm} = u_{nm} + i v_{nm}$ with $u_{nm}$ and $v_{nm}$ sampled independently from the normal distribution with mean zero and variance $1/2$. 

The physical properties of the Rosenzweig-Porter model are determined by the tuning parameter $\gamma$. In the thermodynamic limit $N \to \infty$, one distinguishes between three different phases, which can be characterized by their type of (short-range) level statistics and fractal dimension of the (mid-spectrum) eigenstates. Here, the fractal dimension $d$ is defined in terms of the scaling of the inverse participation ratio $\text{IPR}$ as $\text{IPR} \sim N^{-d}$, where
\begin{equation}
\text{IPR} = \sum_n | \langle n | \psi \rangle|^4,
\end{equation}
with $| \psi \rangle$ denoting the (eigen)state under consideration and the summation running over all basis states $|n \rangle$.

For $0 \le \gamma < 1$, the model is in the ergodic phase. This phase is characterized by Wigner-Dyson level spacing statistics (typically observed for quantum-chaotic systems) and eigenstates with fractal dimension $d = 1$. For $1 < \gamma < 2$, the model is in the fractal phase. This phase is characterized by Wigner-Dyson level spacing statistics and eigenstates with fractal dimension $d = 2 - \gamma$. For $\gamma > 2$, the model is in the localized phase, characterized by Poissonian level statistics (uncorrelated levels, typically observed for integrable systems) and eigenstates with fractal dimension $d = 0$. A numerical investigation of finite-$N$ scalings, which is discussed in some more detail below, can be found in Ref. \cite{Pino19}.

This work focuses on the unitary equivalent of the Rosenzweig-Porter model, which was introduced recently in Ref. \cite{Buijsman22} for the variant with real-valued elements. The eigenvalues of unitary matrices are located on the unit circle in the complex plane, and can thus be parametrized as $\exp( i \, \theta)$ for $\theta \in [-\pi, \pi)$. The density of states for the unitary equivalent of the Rosenzweig-Porter model is uniform, meaning that the spectra can be analyzed without spectral unfolding or the need to take into account edge effects. Constructing a unitary equivalent of the Rosenzweig-Porter model is less trivial than it might seem on a first sight. For example, for unitary matrices $e^{i H_0} \, e^{i M / \sqrt{N^\gamma}}$, the effective Hamiltonian as obtained from the Baker-Campbell-Hausdorff relation is characterized by correlations between the off-diagonal matrix elements that are not present in the Hermitian Rosenzweig-Porter model.

Samples of the unitary equivalent of the Rosenzweig-Porter model can be obtained through stochastic time-evolution of a time-dependent matrix $U(t)$ which is initialized as
\begin{equation}
U(0) =\diag \left( e^{i \theta_1}, e^{i \theta_2}, \dots, e^{i \theta_N} \right)
\end{equation}
with the phases $\theta_n$ ($n = 1, 2, \dots, N$) sampled independently from the uniform distribution ranging over $[-\pi, \pi)$. The dynamics of this unitary matrix are governed by circular Dyson Brownian motion \cite{Dyson62,Dyson72},
\begin{equation}
U(t + dt) = U(t) \, e^{i \sqrt{dt} M},
\label{eq: U-evolution}
\end{equation}
where $M$ is again a sample from the Gaussian unitary ensemble. This matrix is resampled at each evaluation. The time step $dt$ is required to be small enough such that $e^{i \sqrt{dt} M}$ can be approximated by $1 + i \sqrt{dt} M$, that is, $\sqrt{dt}$ is required to be small compared to the mean level spacing given by $2 \pi / N$, meaning that $dt \sim \mathcal{O}(N^{-2})$. 

For the unitary equivalent of the Rosenzweig-Porter model to result, this stochastic process needs to be evaluated up to time $t = N^{- \gamma}$, which means that the required number of time steps scales as $\mathcal{O}(N^{2 - \gamma})$. When using Gaussian elimination, the computational complexity of evolving over a single time step scales as $\mathcal{O}(N^3)$ (matrix-matrix multiplication), from which it follows that the computational complexity of this algorithm scales as $\mathcal{O}(N^{5 - \gamma})$. The Strassen algorithm for matrix-matrix multiplications can reduce the computational complexity to $\mathcal{O}(N^{4.81 - \gamma})$ \cite{Strassen69}. In principle, a further reduction to a computational complexity of $\mathcal{O}(N^{4.37 - \gamma})$ could be archieved using the most state-of-the-art matrix-matrix multiplication algorithm available \cite{Alman21}.

Numerical sampling from the unitary equivalent of the Rosenzweig-Porter ensemble using the algorithm described above is computationally expensive since it requires many evolutions over time intervals of infinitesimal length. As Eq. \eqref{eq: U-evolution} gives the proper time evolution only up to first order, moreover, the results are subject to a loss of accuracy with progressing time. Indeed, this was the approach used in Reference \cite{Buijsman22}. Ref. \cite{Buijsman23} recently proposed an improved algorithm that is not subject to these restrictions. Let 
\begin{equation}
A = U(0) + \sqrt{dt} X,
\end{equation}
where $X$ is an $N \times N$ matrix with complex-valued elements $X_{nm} = u_{nm} + i v_{nm}$ with $u_{nm}$ and $v_{nm}$ sampled independently from the normal distribution with mean zero and unit variance. One can show that a realization $U$ from the unitary equivalent of the Rosenzweig-Porter ensemble can be obtained from the QR decomposition of $A$ as $U = \Lambda Q$, where
\begin{equation}
A = Q R
\end{equation}
with $Q$ being unitary and $R$ being upper triangular with real-valued diagonal elements. The matrix $\Lambda$, making the QR decomposition unique, is obtained from $R$ as
\begin{equation}
\Lambda = \diag \left( \frac{R_{11}}{| R_{11} |}, \frac{R_{22}}{| R_{22} |}, \dots, \frac{R_{NN}}{| R_{NN} |} \right).
\end{equation}
Within this procedure, the time step can be arbitrarily large. A sample from the unitary equivalent of the Rosenzweig-Porter model can be obtained by setting $dt \to N^{- \gamma}$. The computational complexity of this algorithm scales as $\mathcal{O}(N^3)$ (QR decomposition). In what follows, numerical data is obtained using this procedure. It can be of interest to note that this work is the first application of this algorithm.

\section{Short-range spectral statistics} \label{sec: short-range}
In this work, short-range level statistics are quantified through the commonly studied average ratio of consecutive level spacings \cite{Oganesyan07, Atas13}. Let the eigenvalues $\lambda_n$ ($n = 1, 2, \dots, N$) of the unitary matrix that is studied be parametrized as $\lambda_n = \exp(i \theta_n)$ with $\theta_n \in [-\pi, \pi)$, and sorted such that $\theta_1 \le \theta_2 \le \dots \le \theta_N$. The ratios $r_n$ ($n = 1, 2, \dots, N-2$) of consecutive level spacings are then defined as
\begin{equation}
r_n = \min \left( \frac{\theta_{n+2} - \theta_{n+1}}{\theta_{n+1} - \theta_n}, \frac{\theta_{n+1} - \theta_n}{\theta_{n+2} - \theta_{n+1}} \right).
\end{equation}
By construction, $r_n \in [0,1]$. Poissonian and Wigner-Dyson level statistics are characterized by an average value $\langle r \rangle$ given by $\langle r \rangle = 2 \ln(2) - 1 \approx 0.386$ and $\langle r \rangle \approx 0.600$, respectively.

\begin{figure}[t]
\includegraphics[width = 0.95\columnwidth]{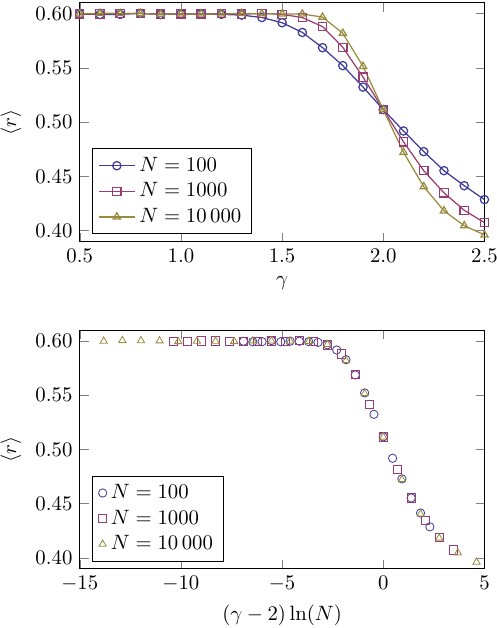}
\caption{The average ratio of consecutive level spacings $\langle r \rangle$ taken over full spectra for dimensions $N = 100$, $N = 1000$, and $N = 10 \, 000$ as a function of $\gamma$ (upper panel) and $(\gamma - \gamma_c) \ln(N)^{1 / \nu}$ with $\gamma_c = 2$ and $\nu = 1$ (lower panel).}
\label{fig: rav}
\end{figure}

Figure \ref{fig: rav} shows the average (taken over full spectra and a large number of samples) ratio of consecutive level spacings as a function of $\gamma$ for dimensions $N = 100$, $N = 1000$, and $N = 10 \, 000$ (upper panel). As expected, the curves tend towards a transition from Wigner-Dyson to Poissonian at the transition between the fractal and the localized phases at $\gamma = 2$. No indications for the transition between the ergodic and the fractal phases at $\gamma = 1$ can be observed. Reference \cite{Pino19} established for the (Hermitian) Rosenzweig-Porter model that a collapse of finite-$N$ curves is observed when plotting $\langle r \rangle$ as a function of $(\gamma - \gamma_c) \ln(N)^{1 / \nu}$ with $\gamma_c = 2$ and $\nu = 1$. A similar scaling is shown in the lower panel, where indeed a collapse can be observed. 

\section{Long-range spectral statistics} \label{sec: long-range}
In terms of the parametrization of the eigenvalues introduced above, long-range spectral statistics are conveniently probed by the spectral form factor
\begin{align}
K(t) 
& = \left \langle \frac{1}{N} \sum_{n,m} e^{i (\theta_n - \theta_m) t} \right \rangle \\
& = \left \langle \frac{1}{N} \left | \sum_n e^{i \theta_n t} \right |^2 \right \rangle,
\end{align}
where again $\langle \cdot \rangle$ denotes an ensemble average \cite{Mehta91}. The spectral form factor can be interpreted as the Fourier transform of the two-point spectral correlation function, where $t$ has the interpretation of a time. Since the phases $\theta_n$ can take values ranging from $-\pi$ to $\pi$ only, the time only takes discrete values $t \in \mathbb{Z}$. For random matrices sampled from the circular unitary ensemble, the spectral form factor $K_\text{CUE}$ can be evaluated analytically as
\begin{equation}
K_\text{CUE}(t) = 
\begin{cases}
|t| / N 	& \text{if } |t| \le N, \\
1				& \text{if } |t| > N.
\end{cases}
\label{eq: sff-rmt}
\end{equation}
For unitary matrices with Poissonian level statistics, one easily finds $K(t) = 1$. Spectral unfolding is conventionally applied when considering spectra with a nonuniform density of states, such that the long-range spectral statistics do not depend on the global density of states. As the global density of states for the unitary equivalent of the Rosenzweig-Porter model is uniform for all values of $\gamma$, no unfolding is required.

\begin{figure}[t]
\includegraphics[width = 0.95\columnwidth]{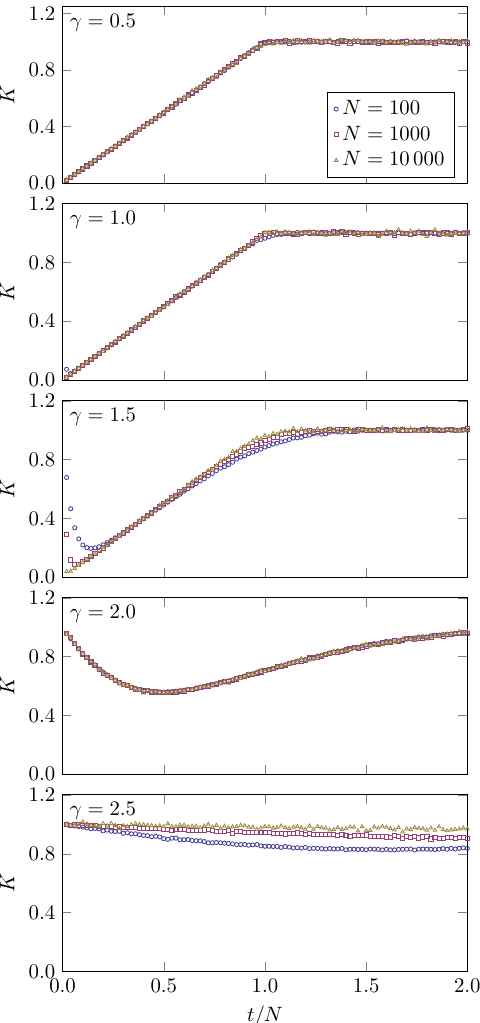}
\caption{The spectral form factor for dimensions $N = 100$, $N = 1000$, and $N = 10 \, 000$ for values of $\gamma$ in each of the ergodic, fractal, and localized phases and at the transitions between them.}
\label{fig: sff}
\end{figure}

Figure \ref{fig: sff} shows the spectral form factor obtained from a large number of spectra as a function of time for values of $\gamma$ in each of the ergodic, fractal, and localized phases for matrix dimensions $N = 100$, $N = 1000$, and $N = 10 \, 000$. In the ergodic phase ($\gamma < 1$), the spectral form factor matches the evaluation for the circular unitary ensemble [Eq. \eqref{eq: sff-rmt}] almost precisely. The fractal phase ($1 < \gamma < 2$) is characterized by intermediate statistics interpolating between the evaluations for the circular unitary ensemble and Poissonian statistics. For $\gamma = 2$, the spectral form factor appears to be scale-invariant in the sense that it is independent of $N$ when considered in terms of the scaled time $t/N$. References \cite{Kunz98, Kravtsov15} report a similar observation for the Hermitian Rosenzweig-Porter model. In the localized phase ($\gamma > 2$), the spectral form factor tends toward $K(t) = 1$ as expected for localized systems with increasing dimension.

The question of interest in this work is whether (long-range) spectral statistics can be used to probe the transition between the ergodic and the fractal phases at $\gamma = 1$. Above, it was illustrated that level spacing (short-range) statistics are insensitive to this transition. Since the spectral form factor can be interpreted as the Fourier transform of the two-point spectral correlation function, one could say that the spectral form factor evaluated at time $t$ probes two-point spectral correlations over a separation $\sim 1/t$. This notion is commonly quantified by the Thoulesss time, after Thouless \cite{Edwards72}. The Thouless time is typically defined as the lowest time from which onwards the spectral form factor matches the evaluation for the (in this case) circular unitary ensemble. See below for the operational definition used in the analysis of the numerical results. The Thouless time has been found useful to quantify the onset of quantum chaos for various physical and random matrix models in recent years \cite{Chan18, Nosaka18, Friedman19, Sierant20-2, Suntajs20, Colmenarez22, Barney23}.

Various quantities referred to as the Thouless time based on, for example, the survival probability \cite{Schiulaz19} or the spreading width of the eigenstates \cite{Kravtsov15} have been established in the literature. The spreading width of the eigenstates gives the energy window over which eigenstates of $H_0$ [or $U(0)$] hybridize due to the off-diagonal term \cite{Kravtsov15, Bogomolny18, Buijsman22}. For the Rosenzweig-Porter model in the fractal or localized phases, it can be obtained from Fermi's golden rule. This width, sometimes referred to as the (inverse) Thouless time, scales as $\mathcal{O}(N^{1 - \gamma})$ in the fractal phase. The spreading width is merely a property of the eigenstates (it can in principle take a different value for each eigenstate), while the spectral form factor is fully obtained from the spectrum. The Thouless time obtained from the spreading width of the eigenstates is thus at most only indirectly related to the Thouless time obtained from the spectral form factor. Nevertheless, a similar scaling is observed below.

A related question is whether one can think of models for which the Thouless time obtained from the spectral form factor on the one hand, and the spreading width of the eigenstates on the other, obey different scalings, which more broadly connects to the timely question on how to detect quantum chaos \cite{Das24}. One could anticipate this to be the case for models with eigenvalue statistics and fractal dimensions of the eigenstates that scale differently with system size. Reference \cite{Das23} recently proposed an experimentally realizable model with this property. The model represents a one-dimensional lattice with random onsite disorder and nearest-neighbor hoppings with a randomized strength. The Hamiltonian has a tuning parameter $\beta > 0$ which parametrizes the spectral correlations, while the fractal dimension of the (midspectrum) eigenstates is determined by $\gamma = - \ln(\beta) / \ln(N)$ with $N$ denoting the matrix dimension. Reminiscent to the Rosenzweig-Porter model, the phase diagram hosts an ergodic ($\gamma < 0$), fractal ($0 < \gamma < 1$), and localized ($\gamma > 1$) phase \cite{Das22}. The spectral statistics are given by those of the Gaussian-$\beta$ random matrix ensemble, meaning that the spectral form factor is independent of $N$ when plotted as a function of $t/N$ (as in the fourth panel of Fig. \ref{fig: sff}) \cite{Dumitriu02}. The Thouless time obtained from the spectral form factor for fixed $\beta$ (let us say $\beta = 0.7$), for this model operationally defined as the lowest time of intersection with the evaluation for the Gaussian orthogonal random matrix ensemble (see the discussion below for details), then obeys $t_\text{Th} \sim N$. As then $\gamma \to 0$ at large matrix dimension, the eigenstates are ergodic (fractal dimension $d=1$), indicating a spreading width that is of the order of the width of the spectrum. This means that $t_\text{Th} \sim 1$ (i.e., it does not scale with $N$) when considering the (inverse) spreading width of the eigenstates as the Thouless time.

The integrated value of the spectral form factor over time is fixed by the presence or absence of level repulsion (see, e.g., Refs. \cite{DeTomasi19, Buijsman20}). Level repulsion is present (absent) if the two-point spectral self-correlation is zero (nonzero). Wigner-Dyson level statistics obey level repulsion, while Poissonian level statistics do not. Specifically,
\begin{equation}
\int_0^\infty \bigg[ 1 - K \bigg(\frac{t}{N} \bigg) \bigg] \, dt = 
\begin{cases}
\pi N & \text{(level repulsion)}, \\
0 & \text{(no level repulsion)},
\end{cases}
\label{eq: sff-int}
\end{equation}
where, for convenience, the sum over discrete times has been written as an integral. Because of this constraint, one cannot trivially define a time from which onwards the spectral form factor matches the evaluation for the circular unitary ensemble. At an operational level, the Thouless time is commonly defined as the time at which the spectral form factor first intersects the evaluation for the corresponding random matrix ensemble, here the circular unitary ensemble (see the references cited above). This seminal operational definition seemingly goes back to Ref. \cite{Suntajs20} on many-body localization.

In numerical studies, the spectral form factor always displays some noise due to averaging over a finite number of samples. For Hermitian models, unfolding inaccuracies typically induce some additional systematic deviations to the spectral form factor. The operational definition of the Thouless time as the lowest time for which the spectral form factor first intersects the spectral form factor for the corresponding random matrix ensemle needs a slight adjustment when these effects are too large. Often, a threshold for the difference between the spectral form factors is introduced. The Thouless time is then operationally defined as the lowest time for which
\begin{equation}
\big| \log K(t) - \log K_\text{CUE}(t) \big| < 10^{- \epsilon},
\end{equation}
for some $\epsilon \simeq 0.4$. Increasing $\epsilon$ lowers the sensitivity of the Thouless time to small deviations between $K_\text{CUE}(t)$ and $K(t)$. In the present study, the number of realizations is large, the spectra do not have edges with possibly deviating statistics as the model is unitary, and no unfolding has to be applied. As such, no threshold for the difference between $K(t)$ and $K_\text{CUE}(t)$ needs to be introduced. Apart from the threshold, the procedure used to determine the Thouless time here is the same as the one used in Ref. \cite{Suntajs20} and the later works.

\begin{figure}[t]
\includegraphics[width = 0.95\columnwidth]{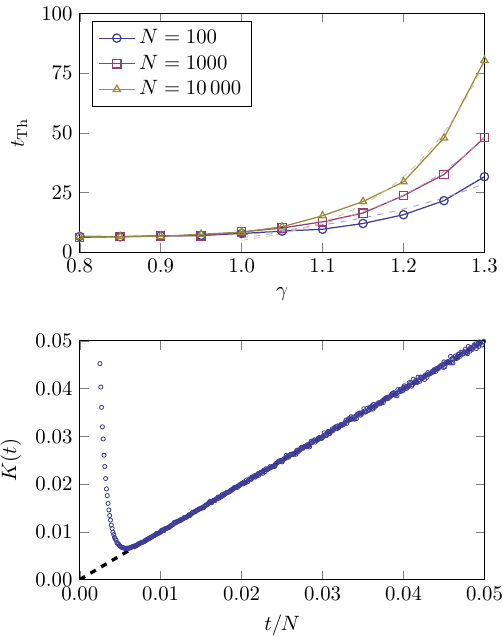}
\caption{Upper panel: The Thouless time $t_\text{Th}$ as obtained from the spectral form factor (see the main text for details) as a function of $\gamma$ around $\gamma = 1$ for dimensions $N = 100$, $N = 1000$, and $N = 10 \, 000$. The dashed lines give fitted (least squares) curves $t_\text{Th} \sim N^{\gamma - 1}$ for $\gamma \ge 1$. Lower panel: An illustration of how the Thouless time is determined. The plot shows the spectral form factor for $N = 10 \, 000$ and $\gamma = 1.3$, combined with the evaluation for the circular unitary ensemble (black dashed curve). The point at which these curves first intersect (here, at $t \approx 80.4$) is marked as the Thouless time.}
\label{fig: time}
\end{figure}

Figure \ref{fig: time} (upper panel) shows the Thouless time $t_\text{Th}$ obtained from the spectral form factor resulting from a large number of spectra as a function of $\gamma$ around $\gamma = 1$ for matrix dimensions $N = 100$, $N = 1000$, and $N = 10 \, 000$. The dashed lines show fitted curves $t_\text{Th} \sim N^{\gamma - 1}$ for $\gamma \ge 1$. Similar to the (inverse) spreading width of the eigenstates introduced above, the Thouless time appears to follow this scaling. This is consistent with the scaling $T_\text{Th} \sim N$ for $\gamma = 2$ that can be read off from Fig. \ref{fig: sff}. The match is not perfect, which might be related to above discussion of the interpretation of the spectral form factor or finite-size effects. For $\gamma < 1$, one observes $t_\text{Th} \sim 1$, indicating the presence of spectral correlations ranging over a finite fraction of the entire spectrum. For clarity, the lower panel shows a graphical illustration of how the Thouless time is obtained. This panel shows the spectral form factor for $N = 10 \, 000$ and $\gamma = 1.3$ combined with the evaluation for the circular unitary ensemble. As can be read off from the upper panel, the points at which the curves first intersect (the Thouless time) is found as $t_\text{Th} \approx 80.4$. Because of the constraint of Eq. \eqref{eq: sff-int}, there is an additional deviation between the curves at $t > t_\text{Th}$ to compensate for the deviation on the interval $t \in [0,t_\text{Th}]$.

For $\gamma < 1$, the Thouless time is of order unity, indicating agreement of the spectral form factor with the random matrix theory expectation almost fully. From $\gamma = 1$ onwards, the Thouless time quickly increases, which indicates a transition to a phase characterized by different long-range spectral two-point correlations. As mentioned before, here a scaling $t_\text{Th} \sim N^{\gamma - 1}$ similar to the (inverse) spreading width of the eigenstates can be observed. Since $\lim_{N \to \infty} t_\text{Th} / N \to 0$ in the fractal phase ($1 < \gamma < 2$), short-range level statistics remain unaffected by the increase of the Thouless time. Consistent results can be observed in Fig. \ref{fig: sff}. These results show that the spectral form factor can be used to probe the transition at $\gamma = 1$ between the ergodic and the fractal phases.

\section{Conclusions and outlook} \label{sec: conclusions}
This work focused on the long-range spectral statistics of the recently introduced unitary equivalent of the Rosenzweig-Porter model. Using the algorithm of Ref. \cite{Buijsman23} to efficiently sample this model, it was argued that the spectral form factor is able to probe the transition between the ergodic and the fractal phases ($\gamma = 1$). More precisely, it was observed that the transition between the ergodic and the fractal phases can be marked as the highest value of $\gamma$ for which the spectral form factor first intersects the spectral form factor of the circular unitary ensemble at times of order unity. This time is often referred to as the Thouless time (see, e.g., Ref. \cite{Suntajs20}), which is a commonly used probe for quantum ergodicity. Similar to the (inverse) spreading width of the eigenstates, the Thouless time is found to scale as $t_\text{Th} \sim N^{\gamma - 1}$ in the fractal phase ($1 < \gamma < 2$). Taking into account that the transition between the fractal and the localized phases can be probed through short-range spectral statistics such as the average ratio of consecutive level spacings, this work establishes that spectral statistics are sufficient to probe both transitions present in the phase diagram.

An aspect arguably worth further investigation is the possible universality of the scale-invariant (in the sense that it is independent of $N$ when considered in terms of the scaled time $t/N$) spectral form factor at the transition between the fractal and the localized phases ($\gamma = 2$), in particular in view of the analytically well-studied two-point spectral correlation function at the transition point \cite{Pandey95, Brezin96, Altland97, Kunz98}. Recently, numerical evidence hinting at such a universality obtained by comparing the long-range spectral statistics for various models at criticality, including the Rosenzweig-Porter model, has been reported \cite{Hopjan23, Hopjan23-2}. Next, the constraint on the spectral form factor imposed by the presence or absence of level repulsion [Eq. \eqref{eq: sff-int}] and the way in which it is taken into account invites for reconsiderations on how to use the spectral form factor as a probe for long-range spectral correlations. In particular, one could ask if the definitions of the Thouless time used here and in the literature could be sharpened.

\begin{acknowledgments}
The author acknowledges support from the Kreitman School of Advanced Graduate Studies at Ben-Gurion University.
\end{acknowledgments}

\bibliography{references}

\end{document}